\documentclass[]{sig-alternate}
\usepackage{etex}
\pdfoutput=1
\AtBeginDocument{

}

\usepackage[utf8]{inputenc}
\usepackage[T1]{fontenc}
\usepackage{listings}
\usepackage{xcolor}
\usepackage{hyperref}
\usepackage{cleveref}
\usepackage{tabularx}
\usepackage{booktabs}
\usepackage{mdframed} 
\usepackage{verbatim} 
\usepackage{paralist}
\usepackage{listings}
\usepackage{courier}
\usepackage{xspace}
\usepackage{balance}
\usepackage{algorithm} 
\usepackage{algorithmic}
\usepackage{multirow}

\newcolumntype{s}{>{\hsize=.75\hsize}X}
\newcolumntype{m}{>{\hsize=1.25\hsize}X}

\newcommand{\tabincell}[2]{\begin{tabular}{@{}#1@{}}#2\end{tabular}}

\makeatletter
\def\@copyrightspace{\relax}
\makeatother

\begin{document}

\title{Test Case Generation for Program Repair:\\ A Study of Feasibility and Effectiveness}

\author{Zhongxing Yu, Matias Martinez, Benjamin Danglot, Thomas Durieux and Martin Monperrus}

\maketitle

\begin{abstract}
Among the many different kinds of program repair techniques, one widely studied family of techniques is called test suite based repair.
Test-suites are in essence input-output specifications and are therefore typically inadequate for completely specifying the expected behavior of the program under repair.
Consequently, the patches generated by test suite based program repair techniques pass the test suite, yet may be incorrect. Patches that are overly specific to the used test suite and fail to generalize to other test cases are called overfitting patches.
In this paper, we investigate the feasibility and effectiveness of test case generation in alleviating the overfitting issue.
We propose two approaches for using test case generation to improve test suite based repair, and perform an extensive evaluation of the effectiveness of the proposed approaches in enabling better test suite based repair on 224 bugs of the Defects4J repository. The results indicate that test case generation can change the resulting patch, but is not effective at turning incorrect patches into correct ones. We identify the problems related with the ineffectiveness, and anticipate that our results and findings will lead to future research to build test-case generation techniques that are tailored to automatic repair systems.
\end{abstract}

\section{Introduction}

Automated program repair holds out the promise of saving debugging cost and patching buggy programs more quickly than humans. Given the great potential, there has been a surge of research on automated program repair in recent years and several different techniques have been proposed \cite{genprog,semfix,nopol,6776507}. These techniques differ in various ways, such as the kinds of used oracles and targeted fault\footnote{In this paper, we use ``fault'' and `bug'' interchangeably} classes \cite{Monperrus2015}.

Among the many different techniques proposed, one widely studied family of techniques is called test suite based repair. Test suite based repair starts with some passing tests as the specification of the expected behavior of the program and at least one failing test as a specification of the bug to be repaired, and aims at generating patches that make all the test cases pass. Based on the used patch generation strategy, test suite based repair can further be informally divided into two general categories, including generate-and-validate techniques and synthesis-based techniques. Generate-and-validate techniques first use certain methods such as genetic programming to generate a set of patches, and then validate the generated patches against the test suite. Representative examples in this category include GenProg \cite{genprog}, PAR \cite{kim2013automatic} and ASTOR \cite{astor2016}. Synthesis-based techniques first use test information to build repair constraint, and then uses a constraint solver such as Z3 to synthesize a patch. Typical examples in this category include SemFix \cite{semfix}, Nopol \cite{nopol}, and Angelix \cite{Mechtaev:2016:ASM:2884781.2884807}. Empirical studies have shown the promise of test suite based repair techniques in tackling the real-life bugs in real-life systems. For instance, GenProg \cite{genprog} and Angelix \cite{Mechtaev:2016:ASM:2884781.2884807} both can generate repairs for large-scale real-world C programs such as wireshark and PHP, while ASTOR \cite{astor2016} and Nopol \cite{nopol} have shown encouraging results \cite{defects4j-repair} on a set of real-life Java programs from the Defects4j benchmark \cite{JustJE2014}.

However, test-suites are in essence input-output specifications and are therefore typically inadequate for completely specifying the expected behavior.
Consequently, the patches generated by test suite based program repair techniques pass the test suite, yet may be incorrect. The patches that are overly specific to the used test suite and fail to generalize to other test cases are called overfitting patches. Overfitting is, indeed, a serious issue associated with test suite based repair techniques and some recent studies have shown that a significant portion of the patches generated by test suite based repair techniques are just overfitting patches \cite{smith2015cure,qi2015efficient,defects4j-repair}. Can test-case generation help in avoiding overfitting?

In this paper, we investigate the feasibility and effectiveness of test case generation in alleviating the overfitting issue. By generating tests using the version-to-be-repaired as an oracle, we obtain tests that would detect errors when the generated patches break some existing functionality. 
The question then arises is how to make best use of the automatically generated test cases.
As we have said, the two categories of tests-suite based repair techniques make use of the tests very differently. By leveraging the specificity of each technique, we design and present two approaches for appropriately using generated tests in the context of program repair: approach MinImpact is appropriate for generate-and-validate techniques that can enumerate patches and approach UnsatGuided is tailored for synthesis-based techniques based on satisfiability.

We evaluate the proposed approaches using 224 bugs of the Defects4J repository \cite{JustJE2014}, which is a database of real-life Java bugs. As jGenProg \cite{astor2016} and Nopol \cite{nopol} are the major open-source test suite based automated repair systems that target Java code, we select them as experimental platforms for generate-and-validate and synthesis-based techniques respectively. With regard to automatic test case generation, we use the state-of-the-art Java unit test generation tool EvoSuite \cite{evosuite}.

By applying program repair augmented with test case generation, our experimental results suggest that:
1) indeed, test case generation influences the output of program repair, it results in different patches;
2) however, it has little impact on the correctness of the final patch, little effect on discarding overfitting patches;
3) one cannot naively feed a repair system with generated tests and thus the careful design of MinImpact and UnsatGuided is required;
4) the overhead can be considered acceptable.

To sum up, the contributions of this paper are: 

\begin{itemize}

\item An analysis of the use of generated test cases in the context of automated program repair, and a characterization of the overfitting problem.

\item Two approaches for using test case generation in test suite based repair. The first one works with generate-and-validate techniques (MinImpact) and the second one for synthesis-based techniques (UnsatGuided).  

\item An extensive evaluation of the effectiveness of the proposed approaches in enabling better test suite based repair on 224 bugs of the Defects4J repository. To the best of our knowledge, this paper is the first systematic and large-scale investigation of the feasibility and effectiveness of test generation for repair.

\end{itemize}

The remainder of this paper is structured as follows. We first present related work in Section 2. Section 3 describes the proposed approaches. Section 4 presents and discusses experimental results, followed by Section 5 which concludes this paper.

\section{Related work}

\subsection{Test Suite Based Program Repair}
Generate-and-validate repair techniques first search within a search space to generate a set of patches, and then validate the generated patches against the test suite. GenProg \cite{genprog}, one of the earliest generate-and-validate techniques, uses genetic programming to search the repair space and generates patches that consist of code snippets copied from elsewhere in the same program.  PAR \cite{kim2013automatic} shares the same search strategy with GenProg but uses 10 specialized patch templates derived from human-written patches to construct the search space. RSRepair \cite{rsrepair} has the same search space as GenProg but uses random search instead, and the empirical evaluation shows that random search can be as effective as genetic programming. AE \cite{6693094} employs a novel deterministic search strategy and uses program equivalence relation to reduce the patch search space. SPR \cite{spr} uses a set of predefined transformation schemas to construct the search space, and patches are generated by instantiating the schemas with condition synthesis techniques. Prophet \cite{prophet} applies probabilistic models of correct code learned from successful human patches to prioritize candidate patches so that the correct patches could have higher rankings. Given that most of the proposed repair systems target only C code, jGenProg and jKali, as implemented in ASTOR \cite{astor2016}, are implementations of GenProg and Kali for Java code. 

Synthesis-based techniques first use the input test suite to extract repair constraints, and then leverage program synthesis to solve the constraint ant get a patch. The patches generated by synthesis-based techniques are generally by design correct with respect to the input test suite.
SemFix \cite{semfix}, the pioneer work in this category of repair techniques, performs controlled symbolic execution on the input test cases to get symbolic constraints, and uses code synthesis to identify a code change that makes all test cases pass. The target repair locations of SemFix are assignments and boolean conditions. To make the generated patches more readable and comprehensible for human beings, DirectFix \cite{directfix} encodes the repair problem into a partial Maximum Satisfiability problem (MaxSAT) and uses a suitably modified Satisfiability Modulo Theory (SMT) solver to get the solution, which is finally converted into the concise patch. Angelix \cite{Mechtaev:2016:ASM:2884781.2884807} uses a lightweight repair constraint representation called “angelic forest” to increase the scalability of DirectFix. Nopol \cite{nopol} uses multiple instrumented test suite executions to synthesize a repair constraint, which is then transformed into a SMT problem and a feasible solution to the problem is finally returned as a patch. Nopol addresses the repair of buggy \emph{if} conditions and missing preconditions in Java code. 

While test suite based techniques are promising, an inherent limitation of them is that the correctness specifications used by
them are the test suites, which are generally available but rarely exhaustive in practice. As a result, the generated patches may just overfit the available test cases, meaning that they will break untested but desired functionality. Several recent studies have shown that overfitting is a serious issue associated with test suite based repair techniques.  Qi et al. \cite{qi2015analysis} find that the vast majority of patches produced by GenProg, RSRepair, and AE avoid bugs simply by functionality deletion. A subsequent study by Smith et al. \cite{smith2015cure} further confirms that the patches generated by of GenProg and RSRepair fail to generalize. The empirical study conducted by Martinez et al. \cite{defects4j-repair} reveals that among the 47 bugs fixed by jGenProg, jKali, and Nopol, only 9 bugs are correctly fixed, the rest being overfitting. 

To improve test suite based repair techniques, we investigate whether the overfitting problem can be alleviated by adding new test cases generated by an automatic test case generation tool.
We expect that the additional tests would make the correctness specification more complete and thus help generate patches that are less likely to be oferfitting and more likely to be correct.

\subsection {Automatic Test Case Generation Tool}

A great deal of research has been targeted on automated test case generation techniques.
In the C realm, DART \cite{godefroid2005dart}, CUTE \cite{sen2005cute}, and KLEE \cite{cadar2008klee} are the three most stable and popular representatives of automatic test case generation tools for C. 
For Java, Randoop \cite{randoop} is the well-known random unit test generation tool. Randoop uses feedback-directed random testing to generate unit tests, and it works by iteratively extending method call sequences with randomly selected method calls and randomly selected arguments from previously constructed sequences. As Randoop test generation process uses a bottom-up approach, it cannot generate tests for a specific class. Other random unit test generation tools for Java include JCrasher \cite{jcrasher}, CarFast \cite{carfast}, T3 \cite{Prasetya2014}. TestFul \cite{testful} and eToc \cite{Tonella:2004:ETC:1013886.1007528}. There are also techniques using various kinds of symbolic execution, such as symbolic PathFinder \cite{Pasareanu:2010:SPS:1858996.1859035} and DSC \cite{Islam:2010:DTC:1868321.1868326}.  
EvoSuite \cite{ESECFSE11} is the state-of-art search-based unit test generation tool for Java and can target a specific class. It uses an evolutionary approach to derive test suites that maximize code coverage, and generates assertions that encode the current behavior of the program. 
To our knowledge, no significant work has been published on the use of test case generation for program repair.

\section{Approach}
\label{sec:approach}

In this section, we present two approaches for improving test suite based repair with test case generation. Given a test suite based repair technique and the initial manually written test suite, our aim is to use additional automatically generated tests to guide the patch generation process towards generating patches that are less likely to be overfitting. The proposed two approaches target generate-and-validate and synthesis-based techniques respectively. 

\subsection{Analysis of the Problem}

\subsubsection{Input domain}
Let us reason about the input space $I$ of a program $P$.
We consider modern and object-oriented programs, where an input point is composed of one or more objects, interacting through a sequence of methods calls. 
In a typical repair scenario, the program is almost correct and thus
a bug only affects the behaviors of a portion of the input domain, which we call the \textbf{buggy input domain} $I_{bug}$.
We call the rest of the input domain, for which the program behaviors are considered as correct $I_{correct}$. 
A patch also affects the behaviors of a portion of the input domain, which we call $I_{patch}$.
For a given bug, a perfect patch impacts behaviors of all input points within $I_{bug}$ and does not change the behaviors for input points within $I_{correct}$, i.e., $I_{bug} = I_{patch}$.

\subsubsection{The bug-exposing test problem}

In regression testing mode, the goal of test case generation is to generate as many new inputs as possible to catch regressions over all the input space. For a certain buggy version, it may generate both input points in $I_{bug}$ and $I_{correct}$. Let us call such a technique $T_{reg}$. 
In regression testing, the behavior of the current program version is used as the oracle. For instance, suppose we have a calculator which incorrectly implements the \texttt{add} function for addition. The code is buggy on the input domain $(10, \_)$ (where \_ means any integer) and is implemented as follows:
\begin{verbatim}
add(x,y) {
  if (x == 10) return x-y;
  else return x+y; }
\end{verbatim}

First, assume that $T_{reg}$ generates a test in the correct input domain $I_{correct}$, for instance on $(5,5)$.
The resulting test, using the existing behavior as oracle would be \texttt{assertEquals(10, add(5,5))}. 

Now, consider what happens when the the generated input lies in $I_{bug}$, say for input pair $(10,8)$.
In this case, $T_{reg}$ would generate test \texttt{assertEquals(2, add(10,8))}.
If the generated inputs lie in $I_{bug}$, the synthesized assertions assert the presence of the actually buggy behavior of the function under test. Put it in another way, the generated assertion encodes the buggy behavior. When the input of a generated test lies in $I_{bug}$, it is called a \textbf{bug-exposing test} in this paper. 

In the context of program repair, it means that some of the generated tests can possibly enforce bad behaviors related with the bug to be repaired. 
Consequently, in case the additional automatically generated tests contain one or several bug-exposing tests and we naively force all generated tests to pass, the patch will necessarily be incorrect.
On the contrary, there will necessarily have failure(s) on bug-exposing tests for a correctly patched program. 

\subsubsection{Overfitting}
\begin{figure}
 \includegraphics[width=\columnwidth]{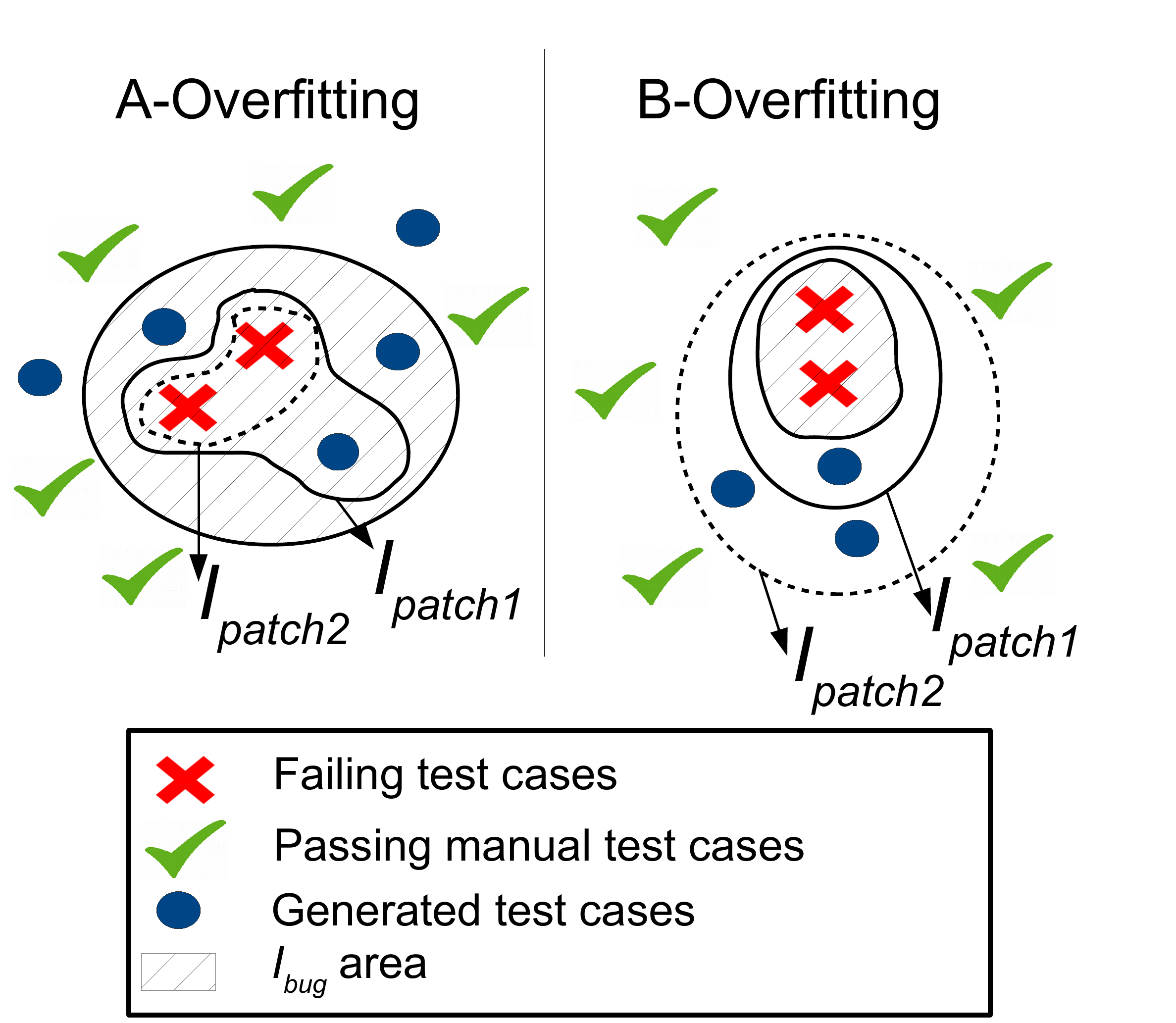}
 \caption{
A-overfitting is a partial patch on the portion of the input domain that is buggy.
B-Overfitting breaks behaviors outside the buggy input domain.
In both cases, a generated test case enables to reason about the difference between $patch_1$ and $patch_2$, and $patch_1$ is better than $patch_2$ in both cases.}
 \label{fig:overfitting}
\end{figure}

Those concepts enable us to define two kinds of overfitting, which are informally presented in \autoref{fig:overfitting}.

\textbf{A-Overfitting}: The overfitting patch only changes the behaviors of some but not all input points within $I_{bug}$, which means that $I_{patch} \subset I_{bug}$. This kind of overfitting patch can be considered as a "partial patch", and encompasses the classical case where there is one single failing test case and the overfitting patch involves specific values related with the failing test case to make it pass.

\textbf{B-Overfitting}: 
The overfitting patch changes the behaviors of some or even all input points within $I_{bug}$, but it also incorrectly breaks the behaviors of some input points within $I_{correct}$.
It means that 
1) the intersection between $I_{correct}$ and $I_{patch}$ is not empty: $I_{correct} \cap I_{patch} \neq \emptyset $
and 2) the intersection between $I_{bug}$ and $I_{patch}$ is also not empty: $I_{bug} \cap I_{patch} \neq \emptyset $

In \autoref{fig:overfitting}, $patch_1$ is better than $patch_2$ because its impact on the input domain is closer to that of the perfect patch.
At the left-hand side, $patch_1$ is less A-overfitting than $patch_2$.
At the right-hand side, $patch_1$ is less B-overfitting than $patch_2$.
In both cases, a generated test case, if exists, enables one to reason about the behavioral difference between $patch_1$ and $patch_2$.

\begin{algorithm}[t]
\begin{algorithmic}[1]
\REQUIRE{A buggy program $P$ and its manually written test suite $TS$}
\REQUIRE{A generate-and-validate repair technique $R_{G\&V}$ and the time budget $TB$}
\REQUIRE{A test case generation tool $T_{reg}$}
\ENSURE{A patch $pt$ to the buggy program $P$}
\STATE{$\{pt_i\} (i=1,2,...,n) \leftarrow R_{G\&V} (P, TS, TB)$}
\IF{$\{pt_i\}(i=1,2,...,n)$ is an empty set}
    \STATE{$pt \leftarrow null$}
\ELSIF{$n==1$}
    \STATE{$pt \leftarrow pt_1$}
\ELSIF{$n>1$}
    \FOR{$i=1$ to $n$}
        \STATE{$\{file_{ij}\}(j=1,2,...,m) \leftarrow getInvolvedFiles(pt_i)$ }
        \STATE{$AGTS_i \leftarrow \emptyset$}
        \FOR{$j=1$ to $m$}
            \STATE{$AGTS_i \leftarrow AGTS_i \cup T_{reg}(P, file_{ij})$}
        \ENDFOR
    
        \STATE{$nbFT_i \leftarrow getNbOfFailingTests(AGTS_i, P, pt_i$)}
    \ENDFOR
    \STATE{$pt \leftarrow pt_1$}
    \STATE{$nbFT \leftarrow nbFT_1$}
    \FOR{$i=1$ to $n$}
        \IF{$nbFT_i < nbFT$}
            \STATE{$pt \leftarrow pt_i$}
            \STATE{$nbFT \leftarrow nbFT_i$}
        \ENDIF
    \ENDFOR
\ENDIF
\RETURN{$pt$}
\end{algorithmic}
\caption{MinImpact: Minimizing the Behavioral Impact for Generate-and-Validate Techniques}
\label{alg:1}
\end{algorithm}

\subsection{MinImpact: Minimizing the Behavioral Impact for Generate-and-Validate Techniques}

Generate-and-validate techniques typically stop searching for other patches when they find the first test-adequate patch (we define test-adequate patch as a patch that passes all of the test cases in the validation test suite). Given that the number of test-adequate patches is generally large \cite{Long:2016:ASS:2884781.2884872} and it is not realistic to return all of them for further check, returning a single patch is a reasonable design choice. However, the first patch is not necessarily the best patch among all of the patches in the search space. It may suffer from overfitting, while other patches discovered later would suffer less from overfitting.
Besides note that even though returning all test-adequate patches for further check is typically unrealistic, generate-and-validate techniques can trivially be configured to search for and list all test-suite adequate patches in the search space.   

Given this situation, we propose MinImpact, which seeks to use additional automatically generated test cases to further check the correctness of a list of test-adequate patches and returns the one with the highest probability of being correct. 
The core idea of MinImpact it to select the test-suite adequate patch that has the smallest behavioral impact. It works as follows. 

For each test-adequate patch established using the initial manually written test suite, we first generate some additional test cases that target the behaviors related with the files involved in the patch. Then, we use these test cases to further validate the correctness of the patch. The intuition is that the more additional test cases fail on a tentatively patched program, the more likely the corresponding patch is an overfitting patch. By comparing the number of test failures on additional test cases for each test-adequate patch, we wish to filter overfitting patches and increase the probability that the returned patch is a correct patch. Note that due to the possible existence of bug-exposing test(s), we cannot simply deem a test-adequate patch as an overfitting patch if the corresponding tentatively patched program fails one or several additional tests. 

Algorithm \autoref{alg:1} describes the proposed approach MinImpact in detail. The algorithm takes as input a buggy program \emph{P} to be repaired, a manually written test suite \emph{TS} which contains some passing tests and at least one failing test, a generate-and-validate technique $R_{G\&V}$, a time budget \emph{TB} allocated for the execution of $R_{G\&V}$, and finally an automatic test case generation tool $T_{reg}$. The output of the algorithm is a patch \emph{pt} to the buggy program \emph{P}. To begin with, the algorithm runs generate-and-validate technique $R_{G\&V}$ against program \emph{P} with the test suite \emph{TS}, and generates a set of test-adequate patches \{$pt_{i}$\} (\emph{i}=1, 2,..., \emph{n}) within the given time budget (line 1). Note that the subscript indicates the order of the patch generated, i.e., $pt_{i}$ is the \emph{i}th patch generated by $R_{G\&V}$. The algorithm directly returns an empty patch if $R_{G\&V}$ generates no patches within the time budget (lines 2-3), and directly returns the generated single patch if $R_{G\&V}$ generates only one patch within the time budget (lines 4-5).  

In case $R_{G\&V}$ generates more than one patch within the time budget, the algorithm then generates additional test cases to further validate the correctness of each patch. For each patch $pt_{i}$, the algorithm first identifies the set of files \{$file_{i,j}$\}($j=1, 2,\ldots, m$) involved in it, i.e., patch $pt_{i}$ makes changes to these files (line 8).
Then, for each file $file_{i,j}$ involved in patch $pt_{i}$, the automatic test case generation tool $T_{reg}$ takes it as input and generates some test cases that target behaviors related with it (line 11, $T_{reg}(P, file_{i,j}$)
). 
The test cases generated for each file of the file set \{$file_{i,j}$\}($j=1, 2,\ldots, m$) are added up to get a test suite $AGTS_i$ which will be used to validate patch $pt_{i}$ (lines 10-12).
Next, the algorithm runs the test suite $AGTS_i$ against the tentatively patched program obtained with patch $pt_{i}$ to see the patch impact, and records the number of failing tests to $nbFT_i$ (line 13). 
When the above process has been completed for each patch, the algorithm selects the patch with the minimum number of failing tests, i.e., $nbFT_i$ ($i=1, 2,\ldots, n$) as the desirable patch (lines 15-22) and returns it (line 24). 
Note if multiple patches have the same number of failing tests and this number is minimum among all of the numbers generated for all of the patches, the algorithm will return the patch generated earliest by $R_{G\&V}$ as the final patch.

\begin{algorithm}[t]
\begin{algorithmic}[1]
\REQUIRE{A buggy program $P$ and its manually written test suite $TS$}
\REQUIRE{A synthesis-based technique $T_{synthesis}$ and the time budget $TB$}
\REQUIRE{A test case generation tool $T_{reg}$}
\ENSURE{A patch $pt$ to the buggy program $P$}
\STATE{$pt_{initial} \leftarrow T_{synthesis}(P, TS, TB)$}
\IF{$pt_{initial} = null$}
    \STATE{$pt \leftarrow null$}
\ELSE
    \STATE{$AGTS \leftarrow \emptyset$}
    \STATE{$pt \leftarrow pt_{initial}$}
    \STATE{$TS_{aug} \leftarrow TS$}
    \STATE{$t_{initial} \leftarrow getPatchGenTime(T_{synthesis}(P, TS, TB))$}
    \STATE{$\{file_{i}\}(i=1,2,...,n) \leftarrow getInvolvedFiles(pt_{initial})$ }
    \FOR{$i=1$ to $n$}
        \STATE{$AGTS \leftarrow AGTS \cup T_{reg}(P, file_{i})$}
    \ENDFOR
    \FOR{$j=1$ to $|AGTS|$}
        \STATE{$t_j \leftarrow AGTS(j)$}
        \STATE{$TS_{aug} \leftarrow TS_{aug} \cup \{t_j\}$}
        \STATE{$pt_{intern} \leftarrow T_{synthesis}(P, TS_{aug}, t_{initial} \times 2)$}
        \IF{$pt_{intern} \neq null$}
            \STATE{$pt \leftarrow pt_{intern}$}
        \ELSE
            \STATE{$TS_{aug} \leftarrow TS_{aug} - \{t_j\} $}
        \ENDIF
    \ENDFOR
\ENDIF
\RETURN{$pt$}
\end{algorithmic}
\caption{UnsatGuided: Incremental Test Suite Augmentation for Synthesis-based Repair}
\label{alg:2}
\end{algorithm}

\subsection{UnsatGuided: Incremental Test Suite Augmentation for Synthesis-based Repair}

Synthesis-based repair techniques such as SemFix and Nopol also suffer from the overfitting issue. The overfitting problem for synthesis-based techniques arises because the repair constraints established using an incomplete test suite may not be strong enough to fully express the intended semantics of a program. Given a certain patch generated by a certain synthesis-based technique, our idea is to generate some additional test cases to strengthen the initial repair constraint.  We wish that a stronger repair constraint will guide synthesis-based techniques towards generating patches that suffer less from overfitting. 

The core problem to handle is the possible existence of bug-exposing test(s) among all of the tests generated.
Because of this, we cannot directly supply all of the generated tests to a synthesis-based technique. The reason is that 
the additional repair constraint enforced by a bug-exposing test can mislead the synthesis process of synthesis-based techniques. Thus, we should make use of the additional constraints enforced by the generated test cases carefully.
 
We now present an approach for using test case generation to improve synthesis-based repair techniques. We call it UnsatGuided. UnsatGuided gradually makes use of the new repair constraints enforced by each generated test case to build a possibly stronger final repair constraint. The key underlying idea is that if the additional repair constraint enforced by a generated test case have contradictions with the repair constraint established using the manually written test suite, then the generated test case is likely to be a bug-exposing test. In this case, the identified bug-exposing test(s) are discarded. 

Algorithm \autoref{alg:2} describes the approach in detail. The input and output of Algorithm \autoref{alg:2} are the same as that of Algorithm \autoref{alg:1} except that the test suite based repair technique used is a synthesis-based technique $T_{synthesis}$. The algorithm directly returns an empty patch if $T_{synthesis}$ generates no patches within the time budget (lines 2-3). In case $T_{synthesis}$ generates an initial patch $pt_{initial}$ within the time budget, the algorithm first conducts a set of initialization steps. It sets the automatically generated test suite \emph{AGTS} to be an empty set temporally (line 5), sets the returned patch \emph{pt} to be the initial patch $pt_{initial}$ temporally (line 6), sets the augmented test suite $TS_{aug}$ to be the manually written test suite \emph{TS} temporally (line 7), and gets the time used by $T_{synthesis}$ to generate the initial patch $pt_{initial}$ and sets $t_{initial}$ to be the value (line 8). Similar to Algorithm \autoref{alg:1}, Algorithm \autoref{alg:2} then identifies the set of files \{$file_i$\}(\emph{i}=1, 2,..., \emph{n}) involved in the initial patch $pt_{initial}$ (line 9) and uses the automatic test case generation tool $T_{reg}$ to generate a set of test cases that target behaviors related with the identified file set \{$file_i$\}(\emph{i}=1, 2,..., \emph{n}) and add them to the test suite \emph{AGTS} (lines 10-12). 

Next, the algorithm will try to use the test suite \emph{AGTS} to refine the initial patch $pt_{initial}$. For each test case $t_j$ in the test suite \emph{AGTS} (line 14), the algorithm first adds it to the augmented test suite $TS_{aug}$ (line 15) and runs technique $T_{synthesis}$ with test suite $TS_{aug}$ and new time budget 2*$t_{initial}$ against program \emph{P} (line 16). The new time budget is used to quickly identify test cases that can potentially contribute to strengthening the repair constraint, and thus improve the scalability of the approach. Then, if the generated patch $pt_{intern}$ is not an empty patch, the algorithm updates the returned patch \emph{pt} with $pt_{intern}$ (lines 17-18). In other words, the algorithm deems test case $t_j$ as a good test case that can help improve the repair constraint. Otherwise, test case $t_j$ is removed from the augmented test suite $TS_{aug}$ (lines 19-20), which implies that the algorithm regards $t_j$ as a bug-exposing test or the kind of tests that will obviously slow down the repair process. After the above process has been completed for each test case in the test suite \emph{AGTS}, the algorithm finally returns patch \emph{pt} as the desirable patch (line 24). 

Note that the order of trying each test case in the test suite \emph{AGTS} matters. Once a test case is deemed as helpful and added to the augmented test suite $TS_{aug}$ permanently, it may impact the result of subsequent try of other tests. The algorithm currently first uses the size of the identified files involved in the patch to determine the test case generation order. The smaller the size of an identified file, the earlier the test case generation tool $T_{reg}$ will generate test cases for it. Then, the algorithm uses the creation time of generated test files and the order of test cases in a generated test file to prioritize test cases. The earlier a test file is created, the earlier its test(s) will be tried by the algorithm. And if a test file contains multiple tests, the earlier a test appears in the file, the earlier the algorithm will try it. Future work will prioritize generated tests according to their potential to improve the repair constraint. 

\subsection{Analysis of Algorithms}
\label{sec:formal-analysis}

\emph{MinImpact}
MinImpact tries to minimize the behavioral impact of the test-adequate patches.
For the generated tests that are in $I_{correct}$, MinImpact minimizes the undesirable impacts of B-Overfitting on input points within $I_{correct}$ (good property). For the generated tests that are in $I_{bug}$, MinImpact tries to select the patch that fixes least buggy input points for the generated tests (possibly bad property). 
With MinImpact, there can possibly exist a trade-off between defeating undesirable impacts on $I_{correct}$ and encouraging maximal overfitting to some specific input points within $I_{bug}$.
The empirical evaluation presented in \autoref{sec:evaluation} will study the prevalence of these phenomena in practice.

\emph{UnsatGuided}
UnsatGuided tries to make use of information from additional tests to build a stronger repair constraint and thus improves the synthesis process. 
If the initial patch is B-Overfitting and the generated tests are in $I_{correct} \cap I_{patch}$, the synthesis is driven towards a new better solution.
If the initial patch is B-Overfitting and the generated tests are in $I_{correct}$ outside $I_{patch}$ or if the initial patch is A-Overfitting and the generated tests are in $I_{correct}$, the patch can either remain the same or be improved.
UnsatGuided tries to identify and discard bug-exposing tests. If the repair constraint established using the manually written test suite is strong, the additional repair constraint enforced by a bug-exposing test is likely to have contradictions with it and UnsatGuided will successfully identify the bug-exposing test and discard it. Then, in case a synthesis-based technique already finds a correct patch with the manually written test suite, UnsatGuided is not likely to change the correct patch into an incorrect patch. However, if the repair constraint established using the manually written test suite is weak, the additional repair constraint enforced by a bug-exposing test does not necessarily have contradictions with it. In this case, UnsatGuided cannot identify the bug-exposing test and will keep it. Once a bug-exposing test is kept, the synthesis will be driven towards keeping the buggy behavior for it. Thus, UnsatGuided may also suffer from the existence of bug-exposing tests. 

\subsection{Technical Insights}

Since a real-world program typically contains many files and a patch to a bug typically just involves changes to one or several files, it is important to generate test cases that focus on the behaviors related with the highly suspicious files,i.e., files with high probability of being buggy. Test cases that focus on behaviors involved in other less suspicious files are less likely to contribute to the process of guiding better patch generation, and instead will at least increase the cost of test case generation and execution. For the proposed two approaches in this paper, we deem the file(s) involved in the initial patch established using the manually written test suite as highly suspicious file(s), and generate tests accordingly.  

We note that a test generation technique sometimes generates tests that are unstable due to the use of non-deterministic APIs such as date and time of day, and these tests are called flaky tests. Since our approaches assume determinism, these tests should be removed. Similar to the work in \cite{JustJIEHF2014,7372009}, we use the following process to remove the flaky tests if they exist: First, we remove all non-compiling test cases. Then, we execute each compliable test suite on the program to be repaired five times consecutively. If any of these executions revealed unstable tests, we then removed these tests and re-compile and re-execute the test suite. This process was repeated until all remaining tests in the test suite passed five times consecutively.

\section{Experimental Evaluation}
\label{sec:evaluation}

In this section, we present an empirical evaluation of the effectiveness of our proposed approaches in improving test suite based repair. 
In particular, we aim at empirically answering
the following research questions:

\begin{itemize}
\item RQ1: Do our approaches impact test suite based repair? Do they yield changes in the generated patches?

\item RQ2: What impacts do our approaches have on the correctness of the generated patches?

\item RQ3: What is the importance of bug-exposing tests?

\item RQ4: What is the time overhead of our approaches?

\end{itemize}

\subsection{Subjects of Investigation}

\subsubsection{Programs}

We selected Defects4J \cite{JustJE2014}, a known database of real faults from real-world Java programs, as the experimental benchmark. Defects4J has different versions and the latest version of the benchmark contains 395 faults from 6 open source projects. Each fault in Defects4J is accompanied with a test suite which contains at least one test case that exposes the fault (manually written). In addition, Defects4J also provides commands to easily access faulty and fixed program versions for each fault, making it relatively easy to analyze them. Among the 6 projects, Mockito has been configured and added to the Defects4J framework very recently (after we start the study presented in this paper) and artifacts related with Mockito are still not in a very stable phase. Thus we do not include the 38 faults for Mockito in our study. Besides, we also discard the 133 faults for Closure compiler as the test cases are organized using scripts rather than the standard JUnit tests, which prevents these tests from running within our repair infrastructure. Consequently, we use the 224 faults of the remaining 4 projects in our experimental evaluation. \autoref{tab:dataset} gives basic information about these 4 subjects. Note that the numbers of lines of source and test code are extracted from the most recent version of each project.

\begin{table}
  \caption{Descriptive Statistics of the 224 Subject Faults}
  \label{tab:dataset}
  \centering
  \small
  \begin{tabular}{|l|r|r|r|r|r|}
    \hline
    Subjects & \#Bugs & \tabincell{c}{Source\\KLoC} & \tabincell{c}{Test\\KLoC} & \#Tests & Dev years \\
    \hline
    JFreechart   & 26  & 96 & 50 & 2,205 & 10 \\
    Commons Math & 106 & 85 & 19 & 3,602 & 14 \\
    Joda-Time    & 27  & 28 & 53 & 4,130 & 14 \\
    Common Lang  & 65  & 22 &  6 & 2,245 & 15 \\
    \hline
\end{tabular}
\end{table}

\subsubsection{Test Suite Based Repair Systems}

For our approaches to be realized, we need stable automatic repair implementations.
In this study, jGenProg \cite{astor2016} and Nopol \cite{nopol} are used as the representatives for generate-and-validate and synthesis-based techniques respectively. We note that jGenProg and Nopol are \emph{open-source} test suite based repair systems that target modern Java code, while PAR \cite{kim2013automatic} is not available and Jaff \cite{4630793} does not support modern Java versions, as required by the bugs of the Defects4J dataset. In addition, it has been shown that Nopol and jGenProg are effective automated repair systems that can tackle real-life faults in real-world programs \cite{defects4j-repair}.

\subsubsection{Automatic Test Case Generation Tool}
  
The automatic test case generation tool used in this study is EvoSuite \cite{evosuite}. To our knowledge, EvoSuite is the state-of-art Java unit test generation tool.
It has been demonstrated as effective in finding potential bugs in open-source and industrial systems and is open-source. 
Besides, as shown in algorithms 1 and 2, our approaches require that the automatic test case generation tool is able to target a specific file of the program under repair. EvoSuite is indeed capable of generating test cases for a specific class.

\begin{table}[h]
\caption{Experimental Results with jGenProg + MinImpact on the Defects4j Repository. Only the bugs with test-suite adequate patches with plain jGenProg are shown.}
\label{tab:results-jgp}
\scalebox{0.82}
{
\begin{tabular}{|l|c|c|r|c|c|c|r|c|r|}
\hline
\multirow{12}{*}{Bug
 ID }& \multicolumn{2}{c|}{jGenProg}& \multicolumn{6}{c|}{jGenProg + MinImpact} \\
 
 \cline{2-9}
 & \rotatebox{90}{Correctness} & \rotatebox{90}{Time
 (hh:mm)} &  \rotatebox{90}{Avg \#EvoTests} & \rotatebox{90}{Avg \# Fails} & \rotatebox{90}{Time (hh:mm)} & \rotatebox{90}{Avg \# test-suite ad. patches} & \rotatebox{90}{Change
 ratio} & \rotatebox{90}{Correct
 ratio} \\ \hline

Chart1 & NO & 00:09 & 97.6 & 0.2  & 02:00 &4.2  & 18/30 & 0/30 \\ 
Chart3 & NO & 00:01 &  129.2 & 0.1  & 02:00 & 4.1 & 30/30 & 0/30 \\ 
Chart5 & NO & 00:02 &  84.4 & 1.7  & 02:00 & 1.9 & 15/30 & 0/30 \\ 
Chart7 & NO & 00:01 &  67 & 0.1  & 02:00 & 1.9 & 18/30 & 0/30 \\ 
Chart13 & NO & 00:01 &  25.5 & 0.1  & 02:00 &22.2 & 30/30 & 0/30 \\ 
Chart15 & NO & 00:02 &  64.7 & 1.9  & 02:00 &  1.4& 30/30 & 0/30 \\ 
Chart25 & NO & 00:01 & 136.8 & 3.9  & 02:00 & 6.6 & 30/30 & 0/30 \\ 
Math2  & NO & 00:11 &  36 & 0.5  & 02:00 & 0.2 & 30/30 & {0/30} \\ 
Math5  & \bf{YES}& 00:21 & 158 & 2.5  & 02:00 & 0.7 & 8/30 & {22/30} \\ 
Math8  & NO & 00:19 &  19.8 & 0.9  & 02:00 & 2 & 19/30 & 0/30 \\ 
Math28 & NO & 00:13 &   7.2 & 0.0  & 02:00 & 8.2 & 26/30 & 0/30 \\ 
Math40 & NO & 00:18 &  27.7 & 2.4  & 02:00 & 1.1 & 29/30 & 0/30 \\ 
Math49 & NO & 00:08 &  109.2 & 5.2  & 02:00 & 2.1 & 22/30 & 0/30 \\ 
Math50 & \bf{YES}& 00:05 &  36.6 & 0.5  & 02:00 & 11.2 & 13/30 & {17/30} \\ 
Math53 & \bf{YES}& 00:05 & 124.5 & 1.1  & 02:00 & 2.8 & 30/30  &{30/30} \\ 
Math70 & \bf{YES}& 00:01 &  21.3 & 1.9  & 02:00 & 1.9 & 13/30 & 17/30\\ 
Math71 & NO & 00:20 &  22 & 0.0  & 02:00 & 0.7 & 30/30 & 0/30 \\ 
Math73 & \bf{YES}& 00:01 &  34.1 & 5.1  & 02:00 & 15.9 & 25/30 & 05/30 \\ 
Math78 & NO & 00:02 &  22 & 0.0  & 02:00 & 1.7 & 29/30 & 0/30 \\ 
Math80 & NO & 00:01 &  45.9 & 2.1  & 02:00 & 3.9 & 28/30 & 0/30 \\ 
Math81 & NO & 00:01 &  45.4 & 2.8  & 02:00 & 5.8 & 30/30 & 0/30 \\ 
Math82 & NO & 00:06 &  17.2 & 0.0  & 02:00 & 1.0 & 0/30 & 0/30 \\ 
Math84 & NO & 00:17 &   5.6 & 0.0  & 02:00 & 0.6 & 22/30 & 0/30 \\ 
Math85 & NO & 00:05 &  25.8 & 3.2  & 02:00 & 4.3 & 8/30 & 0/30 \\ 
Math95 & NO & 00:00 &  19.3 & 1.8  & 02:00 & 8.8 & 28/30 & 0/30 \\ 
Time4  & NO & 00:01 &  56.6 & 1  & 02:00 & 8.7 & 28/30 & 0/30 \\ 
Time11 & NO & 00:01 &  47.6 & 0.2  & 02:00 & 13.9 & 26/30 & 0/30 \\ 
\hline
\end{tabular}
}

\end{table}

\begin{table*}[htbp]
  \scriptsize
  \caption{Experimental Results with Nopol+UnsatGuided on the Defects4j Repository. Only the bugs with test-suite adequate patches found with plain Nopol are shown.}
  \label{tab:nopol-results}
  \begin{tabular}{|l|c|c|r|r|r|c|r|r||l|c|c|r|r|r|c|r|r|}
    \hline
    \multirow{7}{*}{Bug ID} & \multicolumn{2}{c|}{Nopol} & \multicolumn{6}{c||}{Nopol+UnsatGuided} & 
    \multirow{7}{*}{Bug ID} & \multicolumn{2}{c|}{Nopol} & \multicolumn{6}{c|}{Nopol+UnsatGuided}  \\ 
    \cline{2-9}\cline{11-18}
    & \rotatebox{90}{Correctness} & \rotatebox{90}{Time (hh:mm)} & \rotatebox{90}{Avg \#EvoTests} & \rotatebox{90}{Avg \#Contradiction} & \rotatebox{90}{Avg \#Remove} & \rotatebox{90}{Time (hh:mm)} & \rotatebox{90}{{Change ratio}} & \rotatebox{90}{{Correct ratio}} & 
    & \rotatebox{90}{Correctness} & \rotatebox{90}{Time (hh:mm)} & \rotatebox{90}{Avg \#EvoTests} & \rotatebox{90}{Avg \#Contradiction} & \rotatebox{90}{Avg \#Remove} & \rotatebox{90}{Time (hh:mm)} & \rotatebox{90}{{Change ratio}} & \rotatebox{90}{{Correct ratio}} \\
    \hline
Chart1 & NO & 00:02 & 100.4 & 0.0 & 0.0 & 03:00 & 0/30 & 0/30 & Math42 & NO & 00:04 & 59.0 & 0.0 & 0.1 & 03:54 & 24/30 & 0/30\\
Chart5 & NO & 00:01 & 97.7 & 3.0 & 3.5 & 01:18 & 27/30 & 0/30 & Math50 & NO & 00:11 & 36.9 & 0.0 & 0.8 & 04:36 & 28/30 & 0/30\\
Chart9 & NO & 00:01 & 105.5 & 0.0 & 0.0 & 01:00 & 0/30 & 0/30 & Math57 & NO & 00:03 & 21.7 & 0.0 & 0.0 & 00:48 & 15/30 & 0/30\\
Chart13 & NO & 00:02 & 28.4 & 0.0 & 0.0 & 00:24 & 30/30 & 0/30& Math58 & NO & 00:06 & 7.6 & 0.1 & 0.2 & 00:20 & 2/30 & 0/30\\
Chart15 & NO & 00:04 & 123.7 & 0.2 & 0.2 & 06:48 & 27/30 & 0/30 & Math69 & NO & 00:01 & 29.9 & 1.0 & 1.0 & 00:12 & 30/30 & 0/30\\ 
Chart17 & NO & 00:01 & 108.2 & 0.7 & 0.9 & 00:48 & 0/30 & 0/30& Math71 & NO & 00:01 & 31.7 & 0.2 & 0.6 & 00:24 & 25/30 & 0/30 \\
Chart21 & NO & 00:01 & 52.8 & 0.0 & 0.0 & 00:48 & 30/30 & 0/30 & Math73 & NO & 00:01 & 34.5 & 0.1 & 0.3 & 00:18 & 25/30 & 0/30\\ 
Chart25 & NO & 00:01 & 14.7 & 0.0 & 0.0 & 00:12 & 8/30 & 0/30 & Math78 & NO & 00:01 & 33.8 & 1.6 & 1.6 & 00:24 & 28/30 & 0/30\\
Chart26 & NO & 00:03 & 243.2 & 0.4 & 0.6 & 13:36 & 10/10 & 0/10 & Math80 & NO & 00:01 & 45.2 & 0.7 & 1.0 & 00:54 & 29/30 & 0/30\\
Lang44 & \bf{YES} & 00:01 & 101.3 & 0.4 & 0.4 & 00:48 & 3/30 & 30/30 & Math81 & NO & 00:01 & 44.0 & 1.0 & 1.0 & 00:24 & 23/30 & 0/30 \\
Lang51 & NO & 00:01 & 124.0 & 0.5 & 0.5 & 01:00 & 29/30 & 0/30 & Math82 & NO & 00:01 & 17.0 & 0.0 & 0.0 & 00:08 & 0/30 & 0/30\\ 
Lang53 & NO & 00:01 & 97.7 & 0.3 & 0.5 & 00:06 & 26/30 & 0/30 & Math84 & NO & 00:01 & 5.5 & 0.0 & 0.0 & 00:06 & 0/30 & 0/30 \\
Lang55 & \bf{YES} & 00:01 & 20.2 & 0.0 & 0.0 & 00:12 & 30/30 & 30/30 & Math85 & NO & 00:01 & 26.6 & 1.1 & 1.1 & 00:12 & 28/30 & 21/30\\
Lang58 & \bf{YES} & 00:01 & 215.7 & 0.2 & 0.2 & 01:42 & 0/30 & 30/30 & Math87 & NO & 00:01 & 62.2 & 0.0 & 0.0 & 00:54 & 29/30 & 0/30\\
Lang63 & NO & 00:01 & 46.1 & 1.0 & 1.1 & 00:36 & 27/30 & 0/30 & Math88 & NO & 00:01 & 63.0 & 0.0 & 0.0 & 00:30 & 06/30 & 0/30\\
Math7 & NO & 00:16 & 29.2 & 0.0 & 0.0 & 05:00 & 2/30 & 0/30 & Math105 & NO & 00:09 & 45.1 & 0.2 & 0.2 & 04:20 & 29/30 & 0/30\\
Math24 & NO & 00:15 & 132.7 & 0.0 & 2.5 & 24:06 & 10/10 & 0/10 & Time4 & NO & 00:01 & 92.6 & 0.0 & 0.0 & 00:54 & 23/30 & 0/30\\
Math28 & NO & 00:17 & 7.3 & 0.0 & 0.0 & 00:30 & 0/30 & 0/30 & Time7 & NO & 00:01 & 49.7 & 0.4 & 0.4 & 00:54 & 12/30 & 0/30 \\
Math33 & NO & 00:13 & 58.3 & 0.0 & 0.6 & 10:30 & 28/30 & 0/30 & Time11 & NO & 00:04 & 49.9 & 0.0 & 0.0 & 01:36 & 0/30 & 0/30\\
Math40 & NO & 00:16 & 27.7 & 13.0 & 13.0 & 07:00 & 7/30 & 0/30 & Time14 & NO & 00:01 & 22.9 & 0.0 & 0.0 & 00:18 & 24/30 & 0/30\\
Math41 & NO & 00:06 & 40.8 & 1.0 & 1.2 & 02:00 & 27/30 & 0/30 & Time16 & NO & 00:01 & 49.2 & 0.2 & 0.2 & 00:24 & 1/30 & 0/30\\
    \hline
  \end{tabular}
\end{table*}

\subsection{Experimental Setup}

For each of the 224 studied faults in the Defects4J dataset, we run both of our proposed approaches against it. Whenever the test case generation process is invoked, we run EvoSuite 30 times with different seeds to account for the randomness of EvoSuite following the guideline given in \cite{Arcuri:2011:PGU:1985793.1985795}. Since jGenProg is also randomized, we use the same seed for jGenProg. The 30 seeds are 30 integer numbers randomly selected between 1 and 200.

To determine the correctness of the generated patches, we manually analyze them and compare them with the human patches. A generated patch is deemed as correct only if it is exactly the same or semantically equivalent to the human patch. The equivalence is established based on the authors' understanding of the patch. To reduce the possible bias introduced as much as possible, two of the authors analyze the correctness of the patches separately and the results reported in this paper are based on the agreement between them. The time used to analyze each patch differs according to the complexity of it, ranging from several minutes to several hours of work. 

Our experiment is extremely time-consuming. To make the time cost manageable, we set the timeout value, i.e., the input time budget in algorithms 1 and 2, for the approaches MinImpact and UnsatGuided to be 120 and 40 minutes respectively in our experimental evaluation. For combining jGenProg with MinImpact, the experiment was run on Grid’5000, which is a grid for high performance computing \cite{bolze2006grid}. The experiment on combing Nopol with UnsatGuided was run on a cluster consisting of 200 virtual nodes running Ubuntu 16.04 on a single Intel 2.68 GHz Xeon core with 1GB of RAM. As UnsatGuided will try a synthesis-based repair for each test case generated, so the repair process may still cost a lot of time. If so, we reduce the number of considered seeds. This happens for 2 faults (Chart\_26 and Math\_24), for which combining Nopol with UnsatGuided will generally cost more than 13 hours for each EvoSuite seed. Consequently, we run EvoSuite 10 times for these two bugs only for sake of time.

To facilitate future work and better understanding of our work, the experimental materials and results related to this work are available at \cite{replication-site}, including our programs used to run the experiment, test cases generated by EvoSuite, patches generated, and our analysis of the correctness of the patches.

\subsection{Result Presentation}

\autoref{tab:results-jgp} displays our experimental results on combining jGenProg with MinImpact (hereafter referred as jGenProg+\-MinImpact). The \emph{Bug ID} column identifies the buggy version that can be originally repaired by jGenProg. The \emph{Correctness} column and \emph{Time} column under the \emph{jGenProg} column shows the correctness of the original patch generated by jGenProg and the time used to find the patch respectively.  The columns under the column \emph{jGenProg+MinImpact} show the related statistics obtained by running jGenProg+\-Min\-Impact. The \emph{\#Avg \#EvoTests} column shows the number of tests generated by EvoSuite. The \emph{Avg \# Fails} column shows the number of failing EvoSuite tests for the patch returned by jGenProg+MinImpact. The \emph{Time} column shows the time budget used to run approach jGenProg+\-Min\-Impact, which is 2 hours in our experimental evaluation. The \emph{Avg \# test-suite ad. patches} column shows the number of test suite adequate patches found by jGenProg+MinImpact within the time budget. Note that besides column \emph{Time}, the results displayed in the other 3 columns are the average results for the 30 runs using different seeds. Each entry of the column \emph{Change Ratio} is of the form X/Y. Here Y is the number of different seeds used during the experimental process and X is the number of generated patches by jGenProg+MinImpact that are different from the original patch generated by jGenProg among the Y runs. Similarly, each entry of the column \emph{Correct Ratio} is also of the form X/Y. Here Y is also the number of different seeds used during the experimental process and X is the number of generated patches by jGenProg+MinImpact that are correct among the Y runs. 

\autoref{tab:nopol-results} displays our experimental results on combining Nopol with UnsatGuided (hereafter referred as Nopol+\-Unsat\-Guided). Most columns in this table have similar meanings with \autoref{tab:results-jgp} but the repair technique used here is Nopol. Buggy versions that can be originally repaired by Nopol are listed in column \emph{Bug ID}. The \emph{Correctness} and \emph{Time} columns under the column \emph{Nopol} show the correctness of the original patch generated by Nopol and the time used to generate the patch respectively. Related statistics obtained by running Nopol+UnsatGuided are shown in the columns under the column \emph{Nopol+UnsatGuided}. The \emph{\#EvoTests} column shows the number of tests generated by EvoSuite. The \emph{Time} column shows the time used by Nopol+UnsatGuided to calculate the patch. Different from \autoref{tab:results-jgp}, \autoref{tab:nopol-results} have two columns related with removed tests. Note that approach UnsatGuided will remove a generated test if patches cannot be found within the time budget 2*$t_{initial}$, where $t_{initial}$ is the initial time used by Nopol to find a patch, i.e., the items in column \emph{Time} under the column \emph{Nopol}. The number of removed tests are shown in column \emph{Avg \#Remove}. A removed test can be either a test that obviously slows down the repair process or a bug-exposing test. If a test is removed not because of timeout of Nopol, we deem it as a bug-exposing test and the column \emph{Avg \#Contradiction} shows the number of this kind of tests. Similarly, the results displayed in these columns are the average results for the 30 (or 10) runs using different seeds. The last two columns are of the form X/Y and have similar meanings with the last two culumns of \autoref{tab:results-jgp}. Y is the number of different seeds used during the experimental process for both of these two columns, while X refers to the number of generated patches by Nopol+UnsatGuided that are different from the original patch generated by Nopol and the number of generated patches by Nopol+UnsatGuided that are correct resepctively. Note that for the buggy version Chart\_5 and Math\_50, the authors of paper \cite{defects4j-repair} claim that Nopol can generate correct patches for them. However, we carefully study the patches generated by Nopol and the human patches, and think the generated patches are not really correct and they just overfit to the manually written test suite, see our companion website for an explanation \cite{replication-site}. Our project site provides a more deep analysis of this problem. 

Note that even though sometimes the patches generated by running jGenProg+MinImpact or Nopol+UnsatGuided are different from the initial correct patches generated by jGenProg or Nopol, they can also be correct as they are semantically equivalent to the human patches.

\subsection{RQ1: Changed Patches}

We now see whether our approaches can enable test suite based repair techniques to generate different patches.

For generate-and-validate technique jGenProg, we can see from column \emph{Change Ratio} of \autoref{tab:results-jgp} that the proposed approach MinImpact affects the jGenProg process a lot. For the 27 buggy versions that can be initially repaired by jGenProg, the patches generated for 26 buggy versions are changed at least for one seed of EvoSuite after jGenProg+MinImpact is used. If we consider all the different executions (one per seed) over all the buggy versions, this results in 810 patches obtained with jGenProg+MinImpact. Over these 810 patches, 615 patches are different from the original patches generated by using jGenProg only.
We can make similar observations for synthesis-based technique Nopol from the column \emph{Change ratio} of \autoref{tab:nopol-results}. For the 42 buggy versions that can be initially repaired by Nopol, the patches generated for 34 buggy versions are changed at least for one seed of EvoSuite. Likewise, we will get 1220 patches obtained with Nopol+UnsatGuided if we consider all the different executions (one per seed) over all the buggy versions. Over these 1220 patches, 702 patches are different from the original patch generated by running Nopol only. 

Answer for RQ1: The proposed approaches MinImpact and UnsatGuided significantly impact the output of the repair process ({615/810} executions for MinImpact, 636/1220 executions for UnsatGuided).
\emph{It is possible to influence the patches generated by test suite based repair techniques with generated test cases.} 

\subsection{RQ2: Impact on Patch Correctness}

Now, we see whether our proposed approaches impact the correct and incorrect patches established by a basic test suite based technique with only the manually written test suite. In particular, we want to know whether our approaches can change an incorrect overfitting patch into a correct patch. Besides, we also want to observe whether an already correct patch can be unfortunately changed into an overfitting patch instead.

\subsubsection{jGenProg+MinImpact}
For generate-and-validate technique jGenProg, we compare the \emph{Correctness} column under the column \emph{jGenProg} and the \emph{Correct Ratio} column under the column \emph{jGenProg+\-MinImpact} of \autoref{tab:results-jgp} to see the impact of MinImpact on correctness. Even though jGenProg+MinImpact has changed 21 incorrect patches generated by jGenProg at least for some EvoSuite seeds, these newly generated patches are also incorrect. 

It also happens that running jGenProg+MinImpact adversely changes an initial correct patch into an incorrect patch. Take Math\_50 as an example, for which jGenProg successfully repairs the bug as shown in \autoref{tab:results-jgp}. While running jGenProg+MinImpact finds the same correct patch in 17 executions, it finds an incorrect overfitting patch in the other 13 executions and the corresponding tentatively patched program for this patch does not fail any of the generated test cases (note the value in the \emph{\#Avg \#EvoTests} column of \autoref{tab:results-jgp} is the average value over all 30 runs). This incorrect patch indeed has a smaller behavioral impact, and is an example of encouraging maximal overfitting to some points in $I_{bug}$ as discussed in \autoref{sec:formal-analysis}.

Consider another example Math\_70, for which there are on average 1.9 test-adequate patches found, implying that there are commonly 2 test-adequate patches.
One is correct, the other is incorrect.\footnote{The list of test-adequate patches is available on \cite{replication-site}.}
The first patch fails on average 1.1 generated test cases, while the other does on 2.8.  
The second one has a bigger behavioral impact as measured by the generated test cases, because it is B-Overfitting.
This validates our analysis that MinImpact is good at avoiding B-overfitting.

\subsubsection{Nopol+UnsatGuided}

For synthesis-based technique Nopol, we compare the column \emph{Correctness} under the column \emph{Nopol} against the column \emph{Correctness ratio} under the column \emph{Nopol+UnsatGuided} of \autoref{tab:nopol-results} to see the impact of UnsatGuided on correctness. We see that UnsatGuided has successfully helped Nopol to generate one new correct patch. The newly generated correct patch is for the buggy version Math\_85, which contains a bug related with condition. Below is the human patch for it, which has changed the binary relational operator from ">=" into ">".

\begin{verbatim}
1  +  if (fa * fb > 0.0 ) {
2  -  if (fa * fb >= 0.0 ) {
\end{verbatim}

Now consider the patch initially generated by Nopol (without UnsatGuided), it adds a precondition before the buggy statement.

\begin{verbatim}
1  +  if (fa * fb < 0.0 ) {
2     if (fa * fb >= 0.0 ) {
\end{verbatim}

This patch results in a self-contradictory condition, and is thus incorrect. After combining Nopol with UnsatGuided, below is the patch generated. 

\begin{verbatim}
1  +  if (fa * fb != 0.0 ) {
2     if (fa * fb >= 0.0 ) {
\end{verbatim}

This patch is correct as it equates to the human patch semantically, and is found for 21 different seeds of EvoSuite.
For the other 30 incorrect patches intially generated by Nopol, running Nopol+UnsatGuided changes the patch but the resulting new patch is still incorrect. 

For the 3 correct patches generated by Nopol, the generated patches by running Nopol+UnsatGuided are still correct for all seeds of EvoSuite. This justifies the analysis in \autoref{sec:formal-analysis} and is a good attribute of UnsatGuided.

Answer for RQ2: In our experiment, UnsatGuided successfully helps Nopol find a new correct patch. MinImp does not produce any improvement. 
\emph{Overall, the generated test cases have a little positive impact on patch correctness}. This will be further discussed in \autoref{sec:discussion}.
  
\vspace{-0.5em}
\subsection{RQ3: Impact of Bug-exposing Tests}

Is our handling of bug-exposing tests, which is the essence of our approaches, necessary?
We now discuss what will happen if one does not take into account the possible existence of bug-exposing tests. 

For generate-and-validate techniques, the basic usage of generated tests would be to simply deem a patch as an overfitting patch if the tentatively patched program corresponding to the patch fails on some of the generated tests. As shown in \autoref{tab:results-jgp}, all the 30 runs of jGenProg+MinImpact on Math\_53 correctly fix the bug. However, we can see from column \emph{\#Avg \#EvoTests} that the average number of failing EvoSuite tests is 1.1, which means we are very likely to discard the correct patch if we ignore the existence of bug-exposing tests.  

Similarly, one can directly supply generated tests to a synthesis-based technique in case bug-exposing tests are ignored. As shown in \emph{\#Avg \#EvoTests} column of \autoref{tab:nopol-results}, the average number of contradiction tests for the 2 bugs (Lang\_44 and Lang\_58) that can always be correctly repaired by running Nopol+UnsatGuided are not zero. In other words, if we simply supply all of the tests generated by EvoSuite to Nopol, we may get no patches at all for these 2 buggy versions.

Answer for RQ3: Bug-exposing tests have an significant impact on the repair process. 
\emph{Without special consideration for bug-exposing tests, a test suite based repair technique would likely discard correct patches when additional automatically generated tests are used.}

\vspace{-0.5em}
\subsection{RQ4: Time Overhead}

The time cost of an automated repair technique should be manageable for its use in industry. We now discuss the time overhead incurred by our approaches. 

The time cost incurred by approach MinImpact is dominated by the time budget used to calculate a set of test-adequate patches. The budget can vary depending on the computing resources available. For our experiment on combing jGenProg with MinImpact, we set the time budget to be 120 minutes. 

We will mainly see the time overhead incurred by UnsatGuided. For our evaluation, we compare the \emph{Time} column under the column \emph{Nopol} and column \emph{Nopol+UnsatGuided} of \autoref{tab:nopol-results} to see the time overhead incurred by UnsatGuided. First, we see that the approach UnsatGuided incurs some time overhead. Compared with the original repair time used by Nopol to find a patch, the average time used by Nopol+UnsatGuided to get the patch is much longer. Second, the time overhead incurred is acceptable in many cases. For 28 out of 42 buggy versions that can initially be repaired by Nopol, the average repair time used by Nopol+UnsatGuided to get the patch is less than or equal to 1 hour, which is arguably acceptable. Finally, we observe the time overhead incurred by the approach UnsatGuided can be sometimes extremely large. For 3 buggy versions (Chart\_26, Math\_24, and Math\_33), Nopol+UnsatGuided will averagely cost more than 10 hours to get the patch. In particular, the average time used by Nopol+UnsatGuided to get the patch for Math\_24 is 24.1 hours. In those cases, the synthesis process of Nopol is slow and since synthesis is performed for each test case generated, the large amount of time cost is imaginable. To reduce time overhead, future work will explore advanced patch analysis to quickly discard useless tests and thus identify the generated test cases that have the greatest potential to improve the patch.

Answer for RQ4: The time overhead of MinImpact depends on the time budget allocated. The time overhead of UnsatGuided lies under one hour for the majority of the considered faults in our evaluation. 
\emph{However, there is a clear time overhead of using generated test cases on repairing real programs for UnsatGuided.}

\vspace{-0.5em}
\subsection{Discussion}
\label{sec:discussion}

This paper is the first to report empirical facts of the use of generated test cases for automatically repairing real faults in large programs.
On the one hand, we have shown that it is feasible to influence program repair with automatically generated tests.
On the other hand, our results on the Defects4j repository are disappointing: only one new correct patch can be found, which is not particularly effective.

To our understanding, the main explanation is that regression test case generation techniques may generate input points in $I_{bug}$, which are called bug-exposing tests in this paper. In that case, there is no automated oracle to determine whether a generated test lies in $I_{bug}$ or in $I_{correct}$. This is a fundamental conceptual problem, and as we have explained in \autoref{sec:approach}, neither MinImpact nor UnsatGuided can cope with bug-exposing tests well to avoid overfitting to some specific points in $I_{bug}$.

We envision two solutions that can possibly be used to alleviate this problem. First, one can approximate the presence of bug-exposing tests by measuring some kind of distance between the manually written failing test cases and the generated tests. If a generated test is behaviorally close to a failing test case, it is likely to also be in $I_{bug}$. On the contrary, if the distance is far, it is probably in $I_{correct}$.
This calls for future research on behavioral profiling, which is a scarcely researched area.

Second, the other way to combat overfitting lies not in the tests themselves, but in analyzing the patch itself and reasoning $I_{patch}$ directly.
For instance, if $I_{patch}$ is very small, i.e. the patch only operates on a couple of input points, then it is likely to be A-overfitting.
Similarly, if $I_{patch}$ is larger than typical behavioral impacts of bug fixes (e.g, because of functionality removal), the patch is likely to be B-overfitting. Reasoning on $I_{patch}$ directly is the idea behind recent research \cite{prophet,genesis} which tries to make generated patches resemble human patches.

\vspace{-0.5em}
\subsection{Threats to Validity}
In this study, we have used 224 faults of 4 real Java programs from the Defects4J benchmark.
One threat to external validity is that our results would not hold for other benchmarks.  However, Defects4J is the most recent and comprehensive dataset of Java programs currently available, and is developed with the aim of providing real bugs and reproducible studies in software testing research. Besides, Defects4J is extensively used as the evaluation subjects by recent research work in software testing \cite{B.Le:2016:LBF:2931037.2931049,PearsonCJFAEPK2016,Laghari:2016:FSB:2970276.2970308}, and in particular work in automated program repair \cite{defects4j-repair,xiong2016precise}. Another threat to validity is that we evaluate our approaches by considering jGenProg and Nopol as representatives for generate-and-validate and synthesis-based techniques respectively. However, it is not guaranteed that our results would generalize to other test suite based repair systems. jGenProg and Nopol, however, are open-source test suite based repair systems that target modern Java code and can effectively repair real-life faults in real-world programs \cite{defects4j-repair}. A final threat to validity is that we have used just one automatic regression test generation tool, EvoSuite, in the study. However, EvoSuite is arguably the state-of-art Java unit test generation tools, and can target a specific java class as required by our proposed approaches. 

A potential threat to internal validity is that we manually check the patches generated using our proposed approaches and decide whether the generated patches are better compared to the initial patches. We used the human patch as the correctness baseline and the human patch is also used to help us understand the root cause of the bug. This process may introduce errors. To reduce this threat as much as possible, the results reported in this paper are checked and confirmed by two authors of the paper. In addition, the complete results are made available online \cite{replication-site} to let readers gain a more deep understanding of our study and analysis. 

\vspace{-0.5em}
\section{Conclusion}

The recent advances in test suite based program repair have made the once-futuristic idea of repairing programs automatically a achievable reality. 
However, as tests suites are incomplete in nature, using test suites as correctness specification makes test suite based repair techniques suffer from the overfitting issue. 
In this paper, we investigate the feasibility and effectiveness of test case generation in alleviating the overfitting
issue. We have proposed two approaches for using test case generation to improve test suite based repair techniques, and have conducted a large scale experiment to evaluate the proposed approaches. 
The results indicate that test case generation can change the resulting patch, but is quite ineffective at turning incorrect patches into correct ones.  
We anticipate that our results and findings will lead to future research to build test-case generation techniques that are tailored to automatic repair systems.

\newpage
\bibliographystyle{abbrv}
\balance
\bibliography{references} 

\end{document}